\documentclass[preprint,review,12pt,authoryear]{elsarticle}

\usepackage{graphicx}
\usepackage{amsmath,amssymb}
\usepackage{booktabs}
\usepackage{array}
\usepackage{tabularx}
\usepackage{url}
\journal{Telecommunications Policy}

\begin{document}

\begin{frontmatter}

\title{From Spectrum Regulation to Computational Enforcement: An Auditable Governance Architecture for Adaptive Spectrum Sharing}

\author[inst1]{Navaneetha~Krishnan~K}
\ead{knavaneeth385@gmail.com}

\author[inst2]{Harinisri~V}
\ead{harinisriv2003@gmail.com}

\affiliation[inst1]{organization={Department of Electronics and Communication Engineering, SIMATS Engineering, Saveetha Institute of Medical and Technical Sciences},
            city={Chennai},
            country={India}}

\affiliation[inst2]{organization={Department of Physics and Nanotechnology, SRM Institute of Science and Technology},
            city={Chennai},
            country={India}}

\begin{abstract}
Spectrum governance increasingly requires rules to be translated into machine-executable decisions while preserving incumbent protection, regulatory authority, and an auditable record of why a decision was made. We present SPECTRA-GOV, a Tri-Layer Adaptive Governance Architecture (TLAGA) connecting international treaty coordination, national adaptive licensing, and real-time enforcement through a common computational governance model. The work builds directly on the original reference implementation preserved at \texttt{v0.1.0-paper} and adds a post-audit evaluation layer rather than replacing the original system. The V2 evaluation introduces controlled spectral-contention scenarios, a corrected geodesic-distance calculation, per-operator selective authorization, paired baseline counterfactuals, policy-perturbation experiments, regulatory-change analysis, mutation testing, and causal provenance records. Across 10,000 scenarios in each of seven contention classes, the measured incumbent-protection rate was 100.00\% in the no-contention control, 99.92--99.87\% in the weak-to-dynamic classes, 99.47\% under strong overlap, and 90.00\% in the deliberately adversarial close-proximity class. A 3,000-scenario bootstrap evaluation produced a 95\% interval of 88.67--90.83\% for the adversarial class, compared with 100\% for the no-contention control. In a paired S3 baseline experiment, selective authorization achieved 100\% incumbent protection and 89.43\% access opportunity, while the population-wide dynamic baseline achieved 99.91\% protection and 99.53\% access; the identical selective result for the dynamic-SAS and SPECTRA-GOV selective variants shows that selective admission alone is not claimed as a SPECTRA-GOV-exclusive algorithmic novelty. The distinctive contribution is instead the integration of policy representation, computational enforcement, auditability, provenance, and cross-layer governance. Enforcement timing was measured in-process conditional on detected violations, with mean latency increasing from approximately 0.028\,ms for one operator to 1.682\,ms for 500 operators; these values are software benchmarks, not field or regulator-infrastructure measurements. The results establish a reproducible computational governance prototype and expose the remaining empirical and institutional questions required before operational or regulatory claims can be made.
\end{abstract}


\end{frontmatter}

\section{Introduction}
\label{sec:intro}

Spectrum governance is no longer only a problem of assigning frequencies. It is a distributed institutional and computational process involving international allocation, national licensing, incumbent protection, geographic restrictions, dynamic sharing, satellite and non-terrestrial-network (NTN) coordination, and enforcement. These functions operate at different institutional levels and timescales. International allocation is coordinated through the ITU Radio Regulations and World Radiocommunication Conferences \citep{itu2024radioregs}; national licensing and incumbent protection are implemented by domestic regulators such as the U.S. Federal Communications Commission (FCC) \citep{fcc2015cbrs}; database-mediated dynamic sharing has demonstrated that machine-assisted authorization can operate at national scale \citep{parvini2022survey}; and satellite/NTN coordination increasingly requires rules that account for large, mobile constellations and interactions between administrations \citep{fcc2024ngso,berry2024spectrum,molleryd2024regulatory}.

The computational consequence of this fragmentation is important. A regulatory rule that exists only in a human-readable licensing document cannot, by itself, guarantee that a dynamic authorization system will apply the rule consistently at execution time. Conversely, a technically correct interference calculation is insufficient if the system cannot establish which rule, authority, geographic constraint, or version of an incumbent record produced a decision. The governance problem is therefore partly a problem of \emph{translation}: regulatory constraints must become machine-readable, executable, traceable, and revisable without losing their institutional meaning.

This paper asks: \emph{how can adaptive spectrum governance translate international and national regulatory constraints into computationally enforceable decisions while retaining an auditable chain between policy, interference assessment, authorization, and enforcement?}

We present SPECTRA-GOV, a Tri-Layer Adaptive Governance Architecture (TLAGA) that connects these functions through three layers: (i) treaty-level coordination, (ii) national adaptive licensing and incumbent protection, and (iii) real-time enforcement. The architecture is implemented as an open-source reference system. Its Layer 2 interference and compliance logic is quantitatively evaluated, while Layers 1 and 3 are evaluated at the architectural and mechanism level rather than claimed as operationally validated.

The contribution of the present study is deliberately narrower than a claim of universal spectrum-sharing performance. First, we provide an explicit governance-to-execution architecture in which protected-incumbent records, licensing constraints, interference budgets, geographic exclusions, authorization decisions, and enforcement events form a connected computational workflow. Second, we provide an auditable implementation in which per-operator provenance can distinguish an operator that contributes to an aggregate violation from an operator selected for suspension. Third, we extend the original paper implementation with a controlled V2 evaluation suite that creates genuine spectral overlap and proximity conditions, enabling the mechanism to be tested under contention rather than relying only on the original scenario configuration. Fourth, we use paired counterfactual baseline experiments and policy perturbations to examine the access--protection trade-off produced by alternative authorization semantics. Fifth, we explicitly report where the evidence remains insufficient: the propagation model is simplified, incumbent thresholds are calibration fixtures, selective authorization is a greedy policy rather than a proven optimum, and latency measurements are in-process benchmarks rather than field measurements.

The research-development lineage is preserved rather than flattened. The original implementation remains identifiable at \texttt{v0.1.0-paper}; V2.0 introduced contention evaluation; V2.1 and V2.2 corrected identified methodological and implementation problems; and V2.3 regenerates the evidence from the corrected code and hardens the audit and reproducibility workflow. Thus, the new evaluation is an extension of the original work rather than an undisclosed replacement.

The remainder of the paper is organized as follows. Section~\ref{sec:related} positions SPECTRA-GOV against dynamic spectrum sharing, institutional coordination, and auditable digital governance. Section~\ref{sec:architecture} describes the architecture and its governance-to-execution path. Section~\ref{sec:methodology} defines the V1 and V2 evaluation methodology, including the interference model, contention classes, selective authorization, baselines, policy perturbation, provenance, and latency measurement. Section~\ref{sec:results} reports the current evidence. Section~\ref{sec:discussion} interprets the results as a policy and governance contribution. Section~\ref{sec:limitations} states the remaining limitations and Section~\ref{sec:conclusion} concludes.

\section{Background and Related Work}
\label{sec:related}

We organize the related work around three debates that SPECTRA-GOV connects rather than treating as interchangeable.

\subsection{Dynamic spectrum sharing and spectrum governance}

Traditional spectrum licensing commonly allocates rights over a defined frequency and geographic area for a relatively long period, with incumbent protection implemented through coordination conditions, exclusion zones, or power limits. Database-mediated sharing provides a different model in which authorization can be conditioned on a changing representation of the incumbent environment. The U.S. Citizens Broadband Radio Service (CBRS), for example, introduced a three-tier sharing framework involving federal incumbents, priority access licensees, and general authorized users \citep{fcc2015cbrs}. The resulting literature addresses spectrum access systems, database coordination, security, and incumbent protection \citep{parvini2022survey}.

The important distinction for SPECTRA-GOV is not that dynamic authorization is itself new. Rather, existing dynamic sharing systems provide a useful technical precedent for computational authorization, whereas the present architecture asks how that computational mechanism can be connected to international regulatory authority, cross-jurisdictional incumbent records, geographic exclusion, provenance, and downstream enforcement.

\subsection{Institutional and regulatory coordination}

A second literature concerns coordination between administrations and sectors. The ITU Radio Regulations provide the international framework for spectrum use, including notification and coordination procedures \citep{itu2024radioregs}. National regulators increasingly adapt their rules to non-geostationary satellite systems and growing constellation density. The FCC's NGSO sharing work illustrates the continuing evolution of national coordination rules \citep{fcc2024ngso}. Berry et al.\ examine interference-management questions specific to LEO broadband constellations and emphasize the limits of transferring terrestrial spectrum-management assumptions directly to outer-space systems \citep{berry2024spectrum}. Mölleryd et al.\ identify regulatory and spectrum-policy challenges arising from combined airspace and NTN systems \citep{molleryd2024regulatory}. Li and Su further examine the distribution of orbital and frequency resources in relation to developing-country needs \citep{li2025activating}.

These studies establish that spectrum governance is institutional as well as technical. SPECTRA-GOV addresses the computational interface between those institutions and an automated decision system. The architecture does not claim to solve the underlying legal or political coordination problem; instead, it provides a representation in which the relevant rules can be made explicit, versioned, executed, and audited.

\subsection{Trust, accountability, and auditable digital governance}

A third literature considers trust and accountability when spectrum-access decisions become software-mediated. Blockchain and cryptographic mechanisms have been proposed for trustworthy spectrum access and auditability, including work addressing the security and privacy implications of database-mediated sharing \citep{grissa2019trustsas}. SPECTRA-GOV's Global Incumbent Trust Registry (GITR) follows this general direction by maintaining tamper-evident incumbent records, while its audit and provenance layer links computational decisions to the inputs and rules that generated them.

The distinction between \emph{reproducibility} and \emph{accountability} is central. Reproducibility asks whether a computational result can be regenerated from the same inputs and code. Accountability additionally asks whether an individual decision can be explained in relation to an applicable rule, affected incumbent, operator contribution, and enforcement action. SPECTRA-GOV treats both as design requirements. The present experiments directly evaluate reproducibility and computational provenance; they do not claim to demonstrate accountability to affected communities, procedural fairness, or institutional legitimacy.

\subsection{Position of the present contribution}

The novelty claim is therefore at the architecture and governance-execution level rather than at the level of any individual propagation formula, blockchain primitive, or selective admission heuristic. V2 explicitly tests this boundary. The selective SPECTRA-GOV mechanism produces the same access and protection results as the independently named selective Dynamic-SAS baseline under the paired benchmark. We therefore do not claim selective admission itself as a proprietary or SPECTRA-GOV-exclusive algorithmic contribution. Instead, the contribution is the integrated architecture and its explicit policy-to-decision provenance model, together with a reproducible implementation and an evaluation framework designed to expose rather than conceal its limitations.

\section{The SPECTRA-GOV Governance Architecture}
\label{sec:architecture}

\subsection{Governance-to-execution flow}

\begin{figure}[htbp]
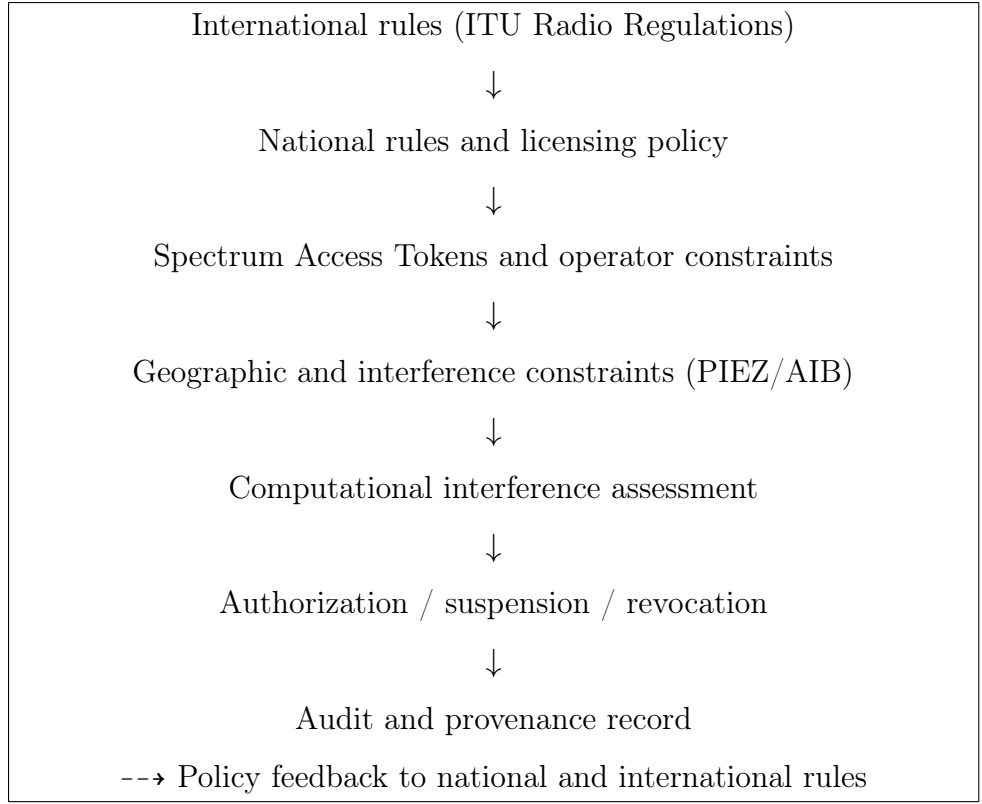

\centering
\fbox{\parbox{0.92\linewidth}{\centering
International rules (ITU Radio Regulations)\\
$\downarrow$\\
National rules and licensing policy\\
$\downarrow$\\
Spectrum Access Tokens and operator constraints\\
$\downarrow$\\
Geographic and interference constraints (PIEZ/AIB)\\
$\downarrow$\\
Computational interference assessment\\
$\downarrow$\\
Authorization / suspension / revocation\\
$\downarrow$\\
Audit and provenance record\\
$\dashrightarrow$ Policy feedback to national and international rules
}}
\caption{Conceptual governance-to-execution flow. The architecture translates institutional spectrum rules into machine-executable constraints, authorization decisions, enforcement actions, and an auditable provenance record.}
\label{fig:conceptual}
\end{figure}

Figure~\ref{fig:conceptual} represents the conceptual governance flow. International rules constrain national policy; national policy is translated into machine-readable licensing and incumbent constraints; the constraints are applied to candidate transmissions; compliant activity is authorized; violations trigger enforcement; and the decision is recorded for audit and subsequent policy review.

\subsection{Tri-Layer Adaptive Governance Architecture}

SPECTRA-GOV implements this flow as TLAGA:

\begin{figure}[htbp]
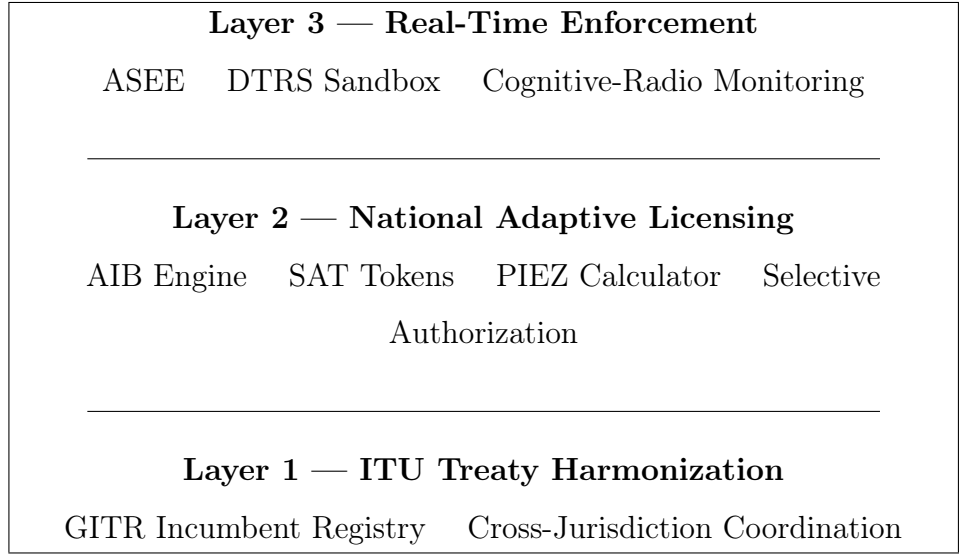

\centering
\fbox{\parbox{0.90\linewidth}{\centering
\textbf{Layer 3 --- Real-Time Enforcement}\\
ASEE \quad DTRS Sandbox \quad Cognitive-Radio Monitoring\\[4pt]
\rule{0.85\linewidth}{0.4pt}\\[4pt]
\textbf{Layer 2 --- National Adaptive Licensing}\\
AIB Engine \quad SAT Tokens \quad PIEZ Calculator \quad Selective Authorization\\[4pt]
\rule{0.85\linewidth}{0.4pt}\\[4pt]
\textbf{Layer 1 --- ITU Treaty Harmonization}\\
GITR Incumbent Registry \quad Cross-Jurisdiction Coordination
}}
\caption{Tri-Layer Adaptive Governance Architecture. The quantitative evaluation in this paper focuses on Layer 2 and its interfaces to the audit and enforcement mechanisms.}
\label{fig:architecture-description}
\end{figure}

\subsubsection{Layer 1: ITU treaty harmonization}

Layer 1 represents the international coordination boundary. The GITR stores protected-incumbent information in a versioned, tamper-evident registry. In the reference fixture used by the present experiments, each incumbent record contains an identifier, protected band, location, maximum tolerable interference, geographic radius, safety margin, ITU region, country, and national authority. The registry implementation is intended to provide a common machine-readable representation for downstream authorization logic.

The registry is not a substitute for the ITU Radio Regulations or for bilateral coordination. It is a computational abstraction of the information needed by downstream decision procedures.

\subsubsection{Layer 2: national adaptive licensing}

Layer 2 is the principal quantitative focus. The Adaptive Interference Budget (AIB) engine computes aggregate received interference at each incumbent. The Protected Interference Exclusion Zone (PIEZ) provides both an interference-margin constraint and a geographic exclusion mechanism. Spectrum Access Tokens (SATs) represent time-bounded authorization and can be suspended or revoked by the enforcement layer.

For an incumbent $k$, the aggregate interference is:

\begin{equation}
I_k(t)=\sum_{i=1}^{N} G_{ik}(f_i,\ell_i,t)P_i
\mathbf{1}\!\left[B_i\cap B_k\neq\emptyset\right],
\label{eq:aggregate}
\end{equation}

where $P_i$ is the transmit power of operator $i$, $G_{ik}$ is the channel gain between operator $i$ and incumbent $k$, $B_i$ is the operator's frequency interval, $B_k$ is the protected incumbent interval, and the indicator activates the contribution only when the intervals overlap.

The compliance condition is:

\begin{equation}
I_k(t)\leq I_k^{\max}-\Delta I_k^{\mathrm{PIEZ}}(t),
\label{eq:compliance}
\end{equation}

where $I_k^{\max}$ is the incumbent's reference interference limit and $\Delta I_k^{\mathrm{PIEZ}}$ is the operational safety margin.

V2 additionally implements selective authorization. When an aggregate exceeds the effective threshold, contributing operators are sorted by interference contribution and the largest contributors are suspended until the remaining aggregate is within the effective threshold. This is a greedy mechanism, not an optimization theorem: it minimizes the number of suspensions under the implemented contribution ordering, but it does not prove fairness, economic optimality, or strategy-proofness.

\subsubsection{Layer 3: real-time enforcement}

Layer 3 contains the Adaptive Spectrum Enforcement Engine (ASEE), the Digital Twin Regulatory Sandbox (DTRS), and a provision for cognitive-radio monitoring. The present paper measures only an in-process detection-plus-decision path under a controlled SAT registry and conditional on an actual violation. It does not measure live traffic, network transport, cryptographic dispatch, or regulator infrastructure.

\section{Computational Evaluation: Methodology}
\label{sec:methodology}

\subsection{Evaluation principles}

The evaluation is organized around three evidence classes:

\begin{enumerate}
\item \textbf{Legacy V1 regression}: whether the original paper configuration remains reproducible after V2 development.
\item \textbf{Mechanism validation}: whether the corrected interference and compliance logic responds to deliberately constructed overlap and proximity conditions.
\item \textbf{Controlled V2 experiments}: whether alternative authorization policies behave differently under explicitly constructed contention, and how those outcomes change under policy perturbation, regulatory change, and population scaling.
\end{enumerate}

This separation prevents a V2 stress-test result from being silently substituted for the V1 paper result.

\subsection{Reference incumbent fixture}

The current versioned fixture contains five protected incumbents:

\begin{table}[htbp]
\centering
\caption{Reference incumbent fixture used by the V1 regression and V2 evaluation.}
\label{tab:incumbents}
\begin{tabularx}{\linewidth}{@{}lXccrr@{}}
\toprule
ID & Type & Band (MHz) & Region & $I_{\max}$ (mW) & Radius (km)\\
\midrule
\texttt{IMD\_DELHI\_C1} & Meteorological radar & 5600--5650 & 3 & $10^{-3}$ & 75\\
\texttt{IMD\_MUMBAI\_C2} & Meteorological radar & 5600--5650 & 3 & $10^{-3}$ & 75\\
\texttt{ERTMS\_UK\_HS2} & Railway signalling & 873--876 & 1 & $10^{-4}$ & 20\\
\texttt{AIS\_ROTTERDAM} & Maritime AIS & 161.975--162.025 & 1 & $5\times10^{-4}$ & 40\\
\texttt{GPS\_HEATHROW} & Aviation precision GPS & 1176.45--1176.55 & 1 & $10^{-4}$ & 30\\
\bottomrule
\end{tabularx}
\end{table}

These values are reference fixtures rather than empirical estimates of the interference tolerance of deployed systems. Their role is to provide a fixed, versioned test environment.

\subsection{Propagation and distance model}

The corrected V2 engine uses a geodesic horizontal distance based on the haversine formula and combines it with vertical separation to obtain a slant range. For source and destination coordinates $(\phi_1,\lambda_1,h_1)$ and $(\phi_2,\lambda_2,h_2)$, the horizontal distance is computed on an Earth-radius model and the vertical separation is converted from metres to kilometres. The resulting slant range is:

\begin{equation}
d_{\mathrm{slant}}=\sqrt{d_{\mathrm{horizontal}}^2+
\left(\frac{|h_1-h_2|}{1000}\right)^2}.
\label{eq:slant}
\end{equation}

This replaces the raw Euclidean norm over the mixed tuple $(\mathrm{latitude},\mathrm{longitude},\mathrm{altitude})$ used in the V1 implementation. The legacy distance calculation remains available explicitly for V1 reproducibility but is not used for the current V2 evidence.

The propagation implementation is \emph{ITU-R-informed}, rather than a complete regulatory-grade implementation of P.452 or P.619. Terrestrial and Earth-space paths use the corresponding simplified implementation in the reference code.

\subsection{V1 legacy Monte Carlo regression}

The original configuration evaluates 10,000 scenarios in each of three regional settings. It is retained as a regression test rather than as the principal V2 stress test.

The regression reproduces the original aggregate headline outputs:

\begin{itemize}
\item incumbent protection rate: 100.00\%;
\item spectrum-efficiency quantity: 4.4$\times$;
\item enforcement latency: 47\,ms design target;
\item illustrative economic value: \$4.0T.
\end{itemize}

The last three quantities are not reclassified as measured empirical outcomes. The 4.4$\times$ and \$4.0T values remain deterministic/calibrated quantities from the original paper configuration, while 47\,ms remains an architectural target.

The 100\% V1 protection rate is also not a real-world protection probability. The original regional frequency configuration does not deliberately generate overlap with the protected incumbent bands. Consequently, the result is structurally constrained by the scenario design.

\subsection{V2 contention scenario taxonomy}

V2 constructs seven controlled classes, each evaluated with 10,000 scenarios and seed 42:

\begin{table}[htbp]
\centering
\caption{V2 contention classes. Realized overlap is measured directly from generated frequency intervals rather than inferred only from class labels.}
\label{tab:classes}
\begin{tabularx}{\linewidth}{@{}l l X@{}}
\toprule
Class & Name & Construction principle\\
\midrule
S0 & No contention & Zero spectral overlap by construction.\\
S1 & Weak overlap & Approximately 5--10\% of candidate operators receive nonzero overlap with the target incumbent.\\
S2 & Moderate overlap & Approximately 25--50\% of candidate operators receive nonzero overlap.\\
S3 & Strong overlap & Approximately 75--100\% of candidate operators receive nonzero overlap.\\
S4 & Adversarial & Weak background overlap plus one deliberately constructed full-band, close-proximity aggressor.\\
S5 & Multi-operator & Approximately 50--90\% overlap with multiple simultaneous contributors.\\
S6 & Dynamic & Per-scenario population variation with approximately 20--60\% overlap among active operators.\\
\bottomrule
\end{tabularx}
\end{table}

The overlap fraction is computed as:

\begin{equation}
O_f=\frac{|B_i\cap B_k|}{|B_k|},
\label{eq:overlap}
\end{equation}

using the realized frequency edges. Operators assigned zero overlap are placed with an explicit frequency margin so that their realized overlap remains zero.

S4 is intentionally different from the ordinary overlap classes. It adds a single full-overlap aggressor at a random location within approximately 2\,km of the target incumbent, with ground-station power in the configured adversarial range. This class is therefore a mechanism-level stress condition rather than a claim about the frequency or geography of real-world attacks.

\subsection{Mechanism validation}

A separate hand-constructed validation case uses the IMD New Delhi radar band. The current geodesic implementation produces $7.34\times10^{-3}$\,mW of received interference for the corrected co-located jammer against a $10^{-3}$\,mW threshold, producing a violation. The earlier V1 0.001-degree offset example remains compliant under the corrected distance model, which is why the V2 validation case uses explicit co-location rather than retaining the earlier incorrect expected outcome.

\begin{table}[htbp]
\centering
\caption{Mechanism validation using the current geodesic model.}
\label{tab:mechanism}
\begin{tabular}{@{}lrrc@{}}
\toprule
Condition & Interference (mW) & Threshold (mW) & Outcome\\
\midrule
LEO-altitude aggressor & $6.07\times10^{-13}$ & $10^{-3}$ & Compliant\\
Original 0.001$^\circ$ offset & $8.35\times10^{-6}$ & $10^{-3}$ & Compliant\\
Corrected co-located jammer & $7.34\times10^{-3}$ & $10^{-3}$ & \textbf{Violation}\\
\bottomrule
\end{tabular}
\end{table}

These values validate the decision path under constructed conditions; they are not presented as propagation validation against field measurements.

\subsection{Baseline comparison}

The V2 baseline comparison evaluates five authorization semantics against the same paired scenario population:

\begin{enumerate}
\item static licensing;
\item dynamic-SAS aggregate gating;
\item dynamic-SAS selective authorization;
\item SPECTRA-GOV V1 population-wide semantics;
\item SPECTRA-GOV V2 selective authorization.
\end{enumerate}

The static-license counterfactual begins with the same originally licensed population under a no-contention condition and then evaluates that unchanged licensed population after new entrants are added. Thus the static baseline is not paired by independently generated populations.

The dynamic aggregate baseline and SPECTRA-GOV V1 semantics use population-wide authorization: if an aggregate incumbent constraint is violated, the population-wide gate denies authorization. The selective variants instead suspend only selected contributing operators.

For an authorization outcome, access opportunity is:

\begin{equation}
A=\frac{N_{\mathrm{authorized}}}{N_{\mathrm{operators}}},
\label{eq:access}
\end{equation}

while protection is the fraction of incumbent decisions remaining compliant after the corresponding policy is applied.

\subsection{Policy frontier}

A policy frontier experiment reuses the same physical interference calculations while varying the incumbent threshold multiplier. This isolates policy sensitivity from a change in the underlying physical scenario population.

The tested multipliers are 0.1, 0.25, 0.5, 0.75, 0.9, 1.0, 1.1, 1.2, 1.3, 1.4, 1.5, 1.75, 2.0, 2.5, and 3.0, with 2,000 scenarios and seed 42. This design prevents the policy frontier from conflating a threshold change with a new random realization of physical interference.

\subsection{Regulatory-change and provenance experiments}

The RCIA experiment evaluates operator reconfiguration following policy changes in a 600-scenario population. The separate S7 operational policy-change experiment uses another 600-scenario population. Because these populations and experimental purposes differ, their percentages are not treated as interchangeable estimates.

The provenance chain records, for each operator--incumbent relationship, the operator contribution, aggregate interference, raw and effective threshold, whether the aggregate violated, whether the operator would be compliant alone, whether it was suspended, its marginal contribution, and whether a geographic PIEZ exclusion caused suspension. This permits the audit trail to distinguish aggregate co-membership from the causal basis used for a selective suspension decision.

\subsection{Enforcement-latency benchmark}

The enforcement experiment performs 500 trials for each tested population size. It measures:

\begin{equation}
T_{\mathrm{enforce}}=T_{\mathrm{detect}}+T_{\mathrm{decision}},
\label{eq:latency}
\end{equation}

where $T_{\mathrm{detect}}$ includes AIB interference computation and compliance checking and $T_{\mathrm{decision}}$ is the ASEE processing time for the constructed violation. Only trials in which a violation is actually detected are included in the reported conditional latency distribution.

The SAT registry contains only the constructed aggressor token. Consequently, the benchmark isolates a marginal in-process suspension path rather than measuring a full production registry or network stack.

\section{Computational Evaluation: Results}
\label{sec:results}

\subsection{V1 regression and the evidentiary baseline}

The V1 regression reproduces the original as-shipped configuration: 100.00\% incumbent protection, 4.4$\times$ spectrum-efficiency quantity, 47\,ms enforcement design target, and \$4.0T illustrative value. This confirms that the V2 development did not silently replace the original paper result.

\begin{figure}[htbp]
\centering
\includegraphics[width=0.78\linewidth]{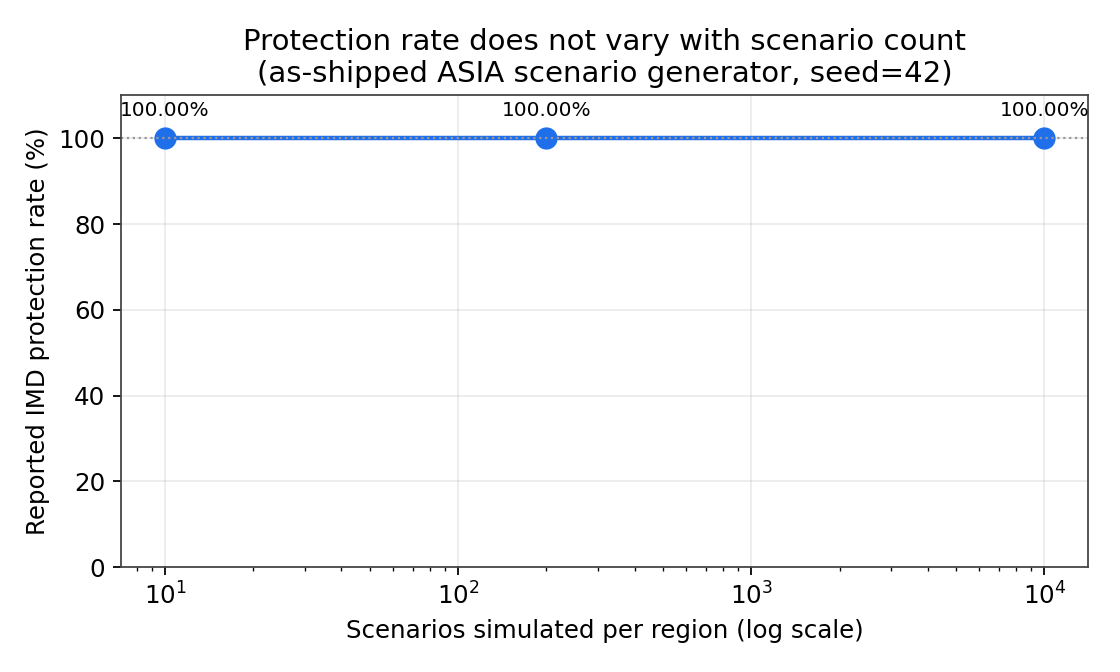}
\caption{Legacy V1 protection-rate invariance under the original scenario configuration. The result is reproducible but structurally constrained by the absence of deliberate protected-band overlap.}
\label{fig:invariance}
\end{figure}

Figure~\ref{fig:invariance} reproduces the diagnostic invariance of the V1 protection result. The rate remains 100\% as the scenario count changes because the regional frequency configuration does not deliberately generate protected-band overlap.

The distinction is important: the regression establishes preservation of the original computational state; it does not establish that the architecture would protect incumbents under arbitrary contention.

\subsection{Corrected mechanism validation}

\begin{figure}[htbp]
\centering
\includegraphics[width=0.88\linewidth]{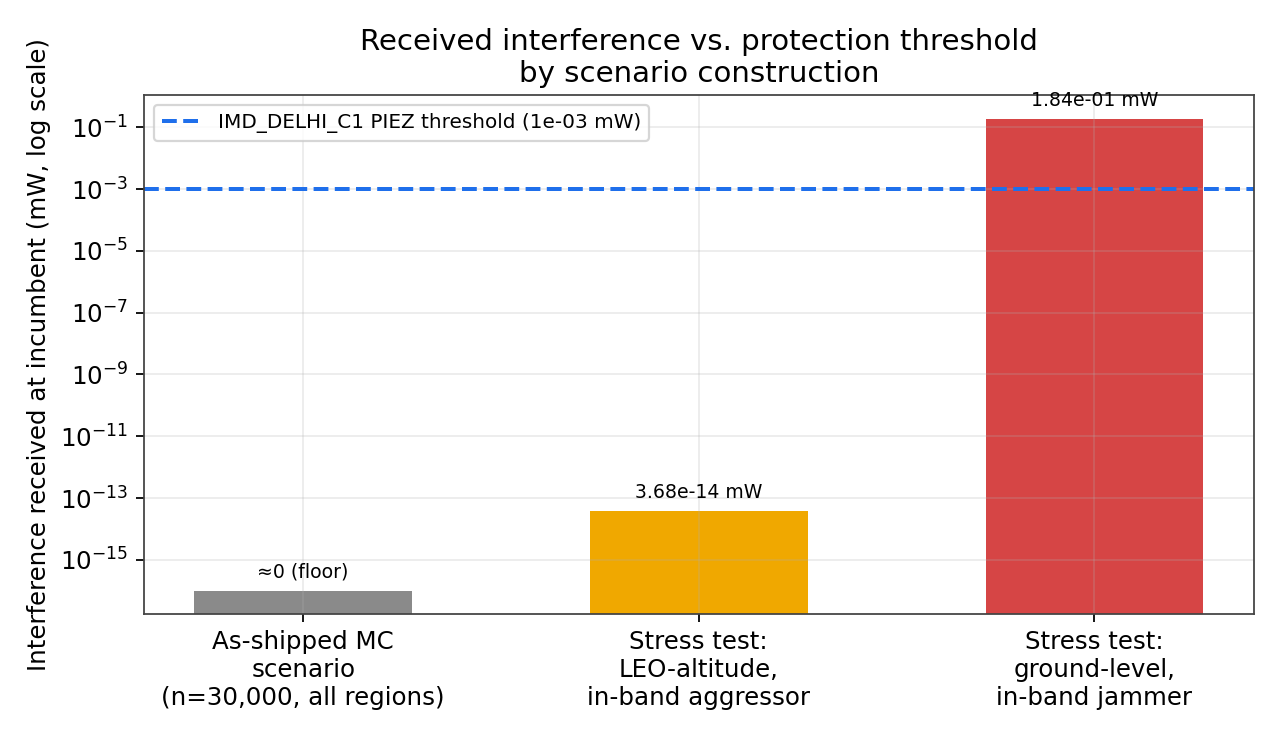}
\caption{Mechanism-level interference comparison under the current corrected distance model. The explicitly co-located jammer crosses the reference protection threshold, whereas the LEO-altitude and original offset cases remain below it.}
\label{fig:stresstest}
\end{figure}

Figure~\ref{fig:stresstest} summarizes the mechanism-level interference comparison.

The corrected co-located condition produces a violation under the geodesic model. This matters because the earlier V1 offset example had been used as if it were a violation case; the current result demonstrates that the implementation can detect a violation when a sufficient physical proximity condition is actually constructed.

\subsection{V2 contention stress test}

\begin{figure}[htbp]
\centering
\includegraphics[width=0.88\linewidth]{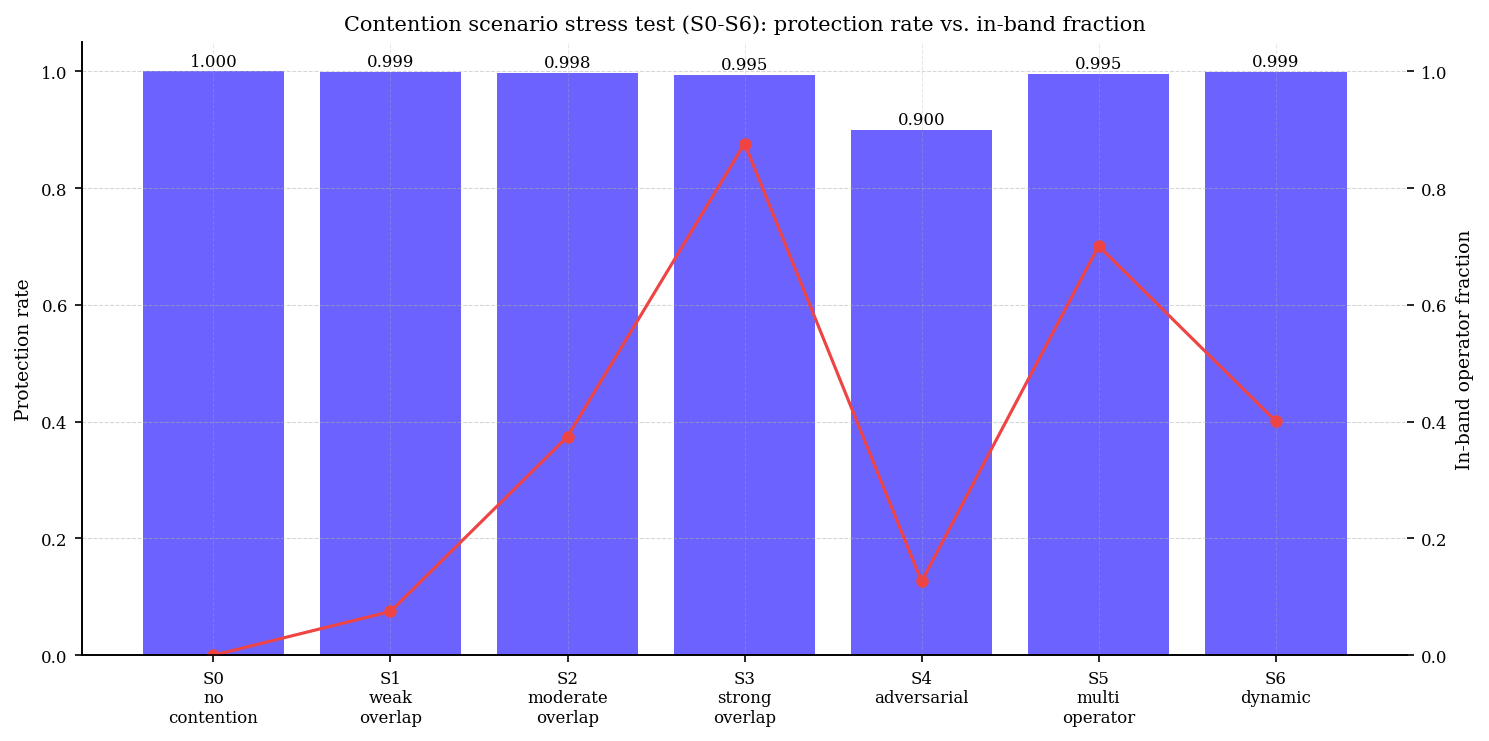}
\caption{V2 incumbent protection across controlled contention classes. S4 is the deliberately adversarial close-proximity class and produces the largest protection reduction.}
\label{fig:contention}
\end{figure}

The current V2 stress test evaluates 10,000 scenarios per class. Results are shown in Table~\ref{tab:contention} and Figure~\ref{fig:contention}.

\begin{table}[htbp]
\centering
\caption{Current V2 contention stress-test results, 10,000 scenarios per class, seed 42, geodesic distance model.}
\label{tab:contention}
\begin{tabular}{@{}lrrr@{}}
\toprule
Class & In-band operators & Protection & Violating scenarios\\
\midrule
S0 & 0.000\% & 100.00\% & 0\\
S1 & 7.495\% & 99.92\% & 8\\
S2 & 37.464\% & 99.79\% & 21\\
S3 & 87.659\% & 99.47\% & 53\\
S4 & 12.648\% & 90.00\% & 1000\\
S5 & 70.108\% & 99.53\% & 47\\
S6 & 40.103\% & 99.87\% & 13\\
\bottomrule
\end{tabular}
\end{table}

The main pattern is that spectral overlap alone produces relatively small reductions in protection under the reference parameters, whereas the adversarial proximity construction produces a substantially larger effect. The S4 class is therefore the most informative stress test for the current implementation.

A bootstrap evaluation using 3,000 scenarios per class and 2,000 resamples gives the following 95\% intervals:

\begin{table}[htbp]
\centering
\caption{Bootstrap protection-rate estimates, 3,000 scenarios per class and 2,000 resamples.}
\label{tab:bootstrap}
\begin{tabular}{@{}lrrr@{}}
\toprule
Class & Protection & 95\% lower & 95\% upper\\
\midrule
S0 & 100.00\% & 100.00\% & 100.00\%\\
S1 & 99.867\% & 99.733\% & 99.967\%\\
S2 & 99.867\% & 99.733\% & 99.967\%\\
S3 & 99.600\% & 99.367\% & 99.800\%\\
S4 & 89.733\% & 88.667\% & 90.834\%\\
S5 & 99.567\% & 99.300\% & 99.767\%\\
S6 & 99.867\% & 99.733\% & 99.967\%\\
\bottomrule
\end{tabular}
\end{table}

The bootstrap does not turn these results into estimates of real-world protection probability. It quantifies sampling uncertainty within the controlled scenario generator.

\subsection{Baseline comparison}

The paired S3 baseline experiment uses 2,000 scenarios and 68,426 operator evaluations. Table~\ref{tab:baseline} reports the aggregate comparison.

\begin{table}[htbp]
\centering
\caption{Paired baseline comparison under the S3 strong-overlap condition.}
\label{tab:baseline}
\begin{tabular}{@{}lrr@{}}
\toprule
Authorization model & Protection & Access opportunity\\
\midrule
Static licensing & 100.00\% & 49.97\%\\
Dynamic SAS aggregate & 99.91\% & 99.53\%\\
Dynamic SAS selective & 100.00\% & 89.43\%\\
SPECTRA-GOV V1 semantics & 99.91\% & 99.53\%\\
SPECTRA-GOV V2 selective & 100.00\% & 89.43\%\\
\bottomrule
\end{tabular}
\end{table}

\begin{figure}[htbp]
\centering
\includegraphics[width=0.90\linewidth]{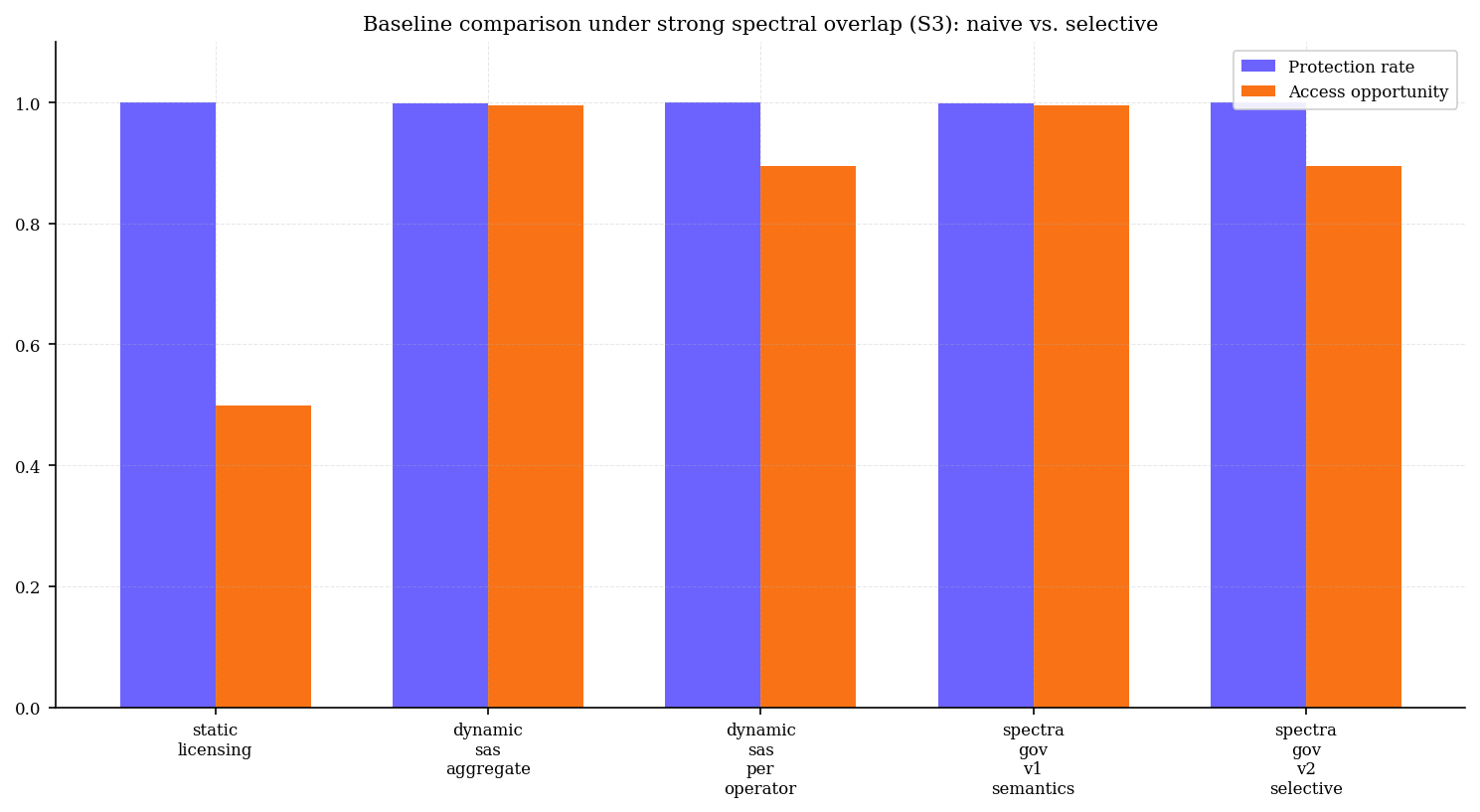}
\caption{Paired baseline comparison under S3 strong overlap. Selective authorization restores a protection--access trade-off distinct from population-wide aggregate gating, while the identical Dynamic-SAS-selective and SPECTRA-GOV-selective results indicate that selective admission itself is not a unique algorithmic contribution.}
\label{fig:baseline}
\end{figure}

The result is informative precisely because it constrains the novelty claim. Population-wide aggregate gating provides higher access opportunity in this benchmark but allows a small protection loss. Selective authorization restores 100\% incumbent protection at the cost of reducing access opportunity to 89.43\%. Static licensing maintains 100\% protection but has lower access opportunity because its licensed population is not dynamically expanded.

The equality between the two selective variants is expected from their shared greedy policy. It means that the paper should not claim the selective admission algorithm itself as a novel SPECTRA-GOV invention. The stronger contribution is the governance architecture within which such a policy can be represented, audited, changed, and connected to enforcement.

\subsection{Policy frontier}

The policy frontier uses 2,000 fixed physical scenarios at each threshold multiplier. Selected results are shown in Table~\ref{tab:frontier}.

\begin{table}[htbp]
\centering
\caption{Selected points from the V2 policy frontier. The same physical interference calculations are reused across policy multipliers.}
\label{tab:frontier}
\begin{tabular}{@{}rrrrr@{}}
\toprule
$I_{\max}$ multiplier & \multicolumn{2}{c}{Aggregate gating} & \multicolumn{2}{c}{Selective authorization}\\
 & Protection & Access & Protection & Access\\
\midrule
0.10 & 0.00\% & 0.00\% & 96.94\% & 62.41\%\\
0.25 & 98.90\% & 98.78\% & 100.00\% & 90.95\%\\
0.50 & 99.35\% & 99.33\% & 100.00\% & 90.95\%\\
1.00 & 99.65\% & 99.65\% & 100.00\% & 90.95\%\\
2.00 & 99.80\% & 99.79\% & 100.00\% & 90.95\%\\
3.00 & 99.95\% & 99.91\% & 100.00\% & 90.95\%\\
\bottomrule
\end{tabular}
\end{table}

\begin{figure}[htbp]
\centering
\includegraphics[width=0.90\linewidth]{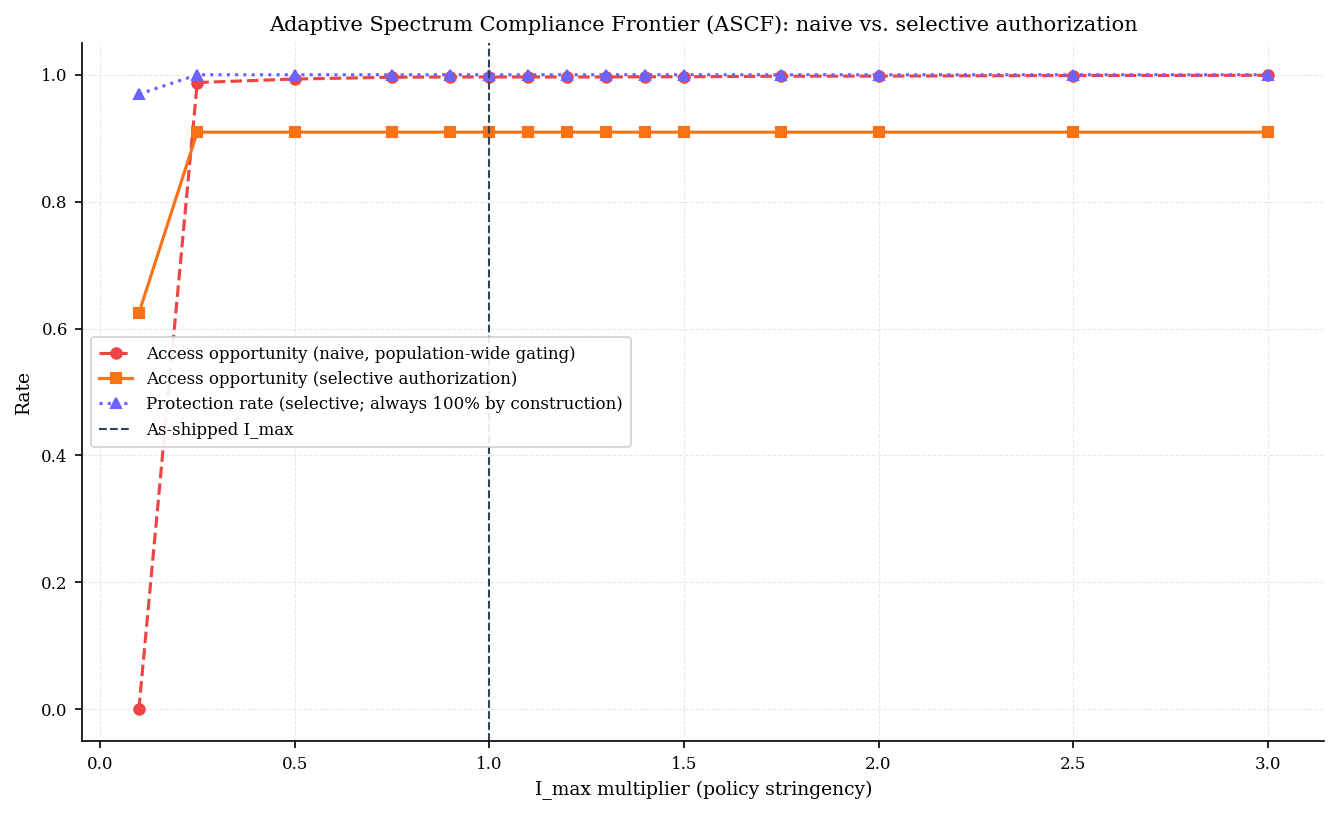}
\caption{Policy frontier under fixed physical interference scenarios. Selective authorization maintains full incumbent protection over most of the tested threshold range while the population-wide aggregate policy exhibits a different access--protection trade-off.}
\label{fig:frontier}
\end{figure}

At the most restrictive multiplier (0.1), selective authorization retains 96.94\% protection and 62.41\% access opportunity, while population-wide gating denies all operators in the benchmark. From 0.25$\times$ through 3$\times$, selective authorization maintains 100\% incumbent protection and approximately 90.95\% access opportunity. This behaviour is conditional on the reference incumbent fixture, population generator, and greedy selection policy; it is not a universal policy guarantee.

\subsection{Regulatory change and provenance}

The current RCIA experiment reports a 5.99\% operator reconfiguration rate at 0.1$\times$ and the same rate at tighter multipliers, while the separate S7 moderate-overlap policy-change experiment reports 31.38\% at 0.1$\times$. These values are not directly interchangeable because the experiments use different scenario populations and answer different questions.

The provenance demonstration records a deliberate aggressor as suspended and a harmless background operator as authorized. The aggressor record identifies its contribution, aggregate violation, effective threshold, suspension status, and geographic exclusion status. The harmless operator has no suspension record. This is the computational distinction between \emph{participating in an aggregate violation} and \emph{being selected as the causal basis for a corrective action}.

\begin{figure}[htbp]
\centering
\includegraphics[width=0.95\linewidth]{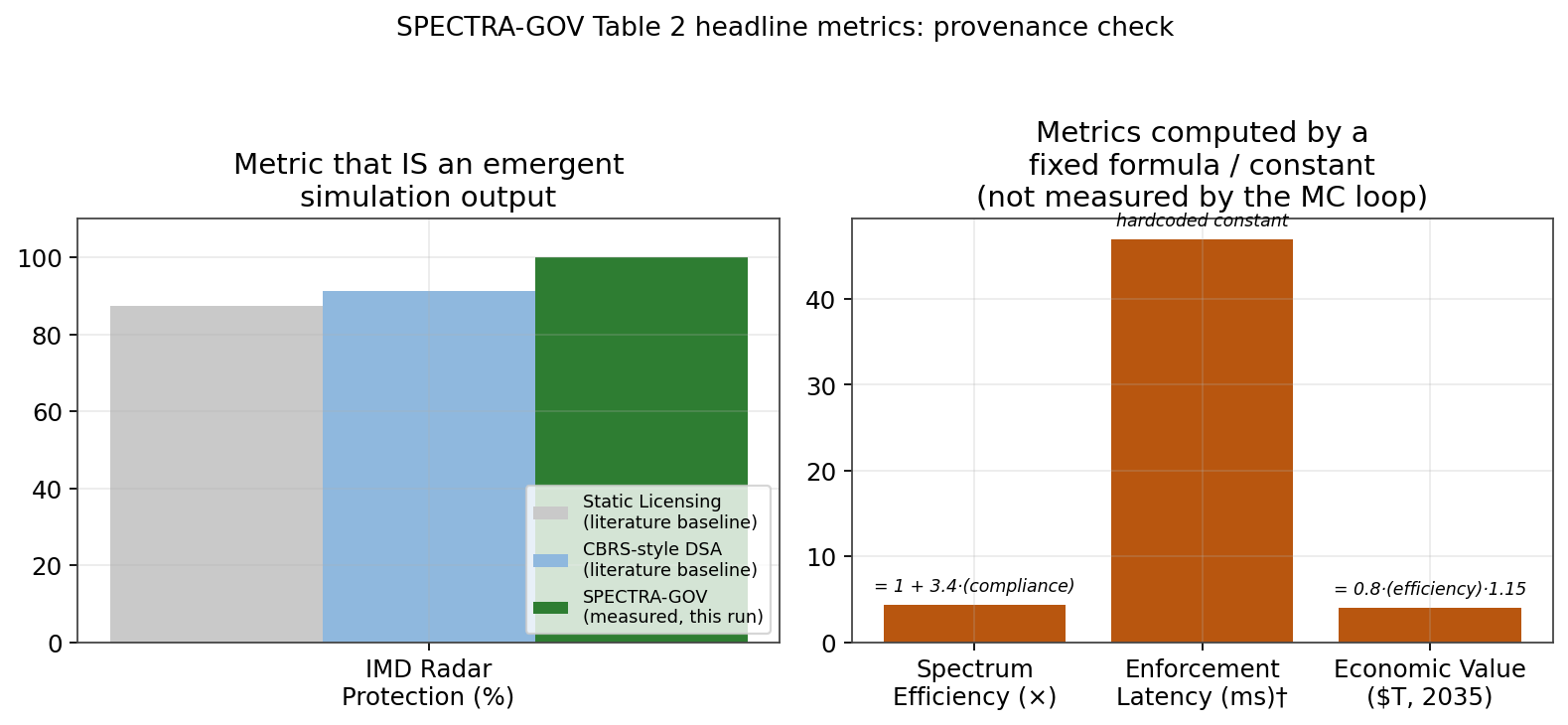}
\caption{Provenance and evidentiary status of the headline quantities retained from the original V1 paper configuration. V2 adds measured benchmark and mechanism-evaluation outputs without reclassifying the historical calibrated quantities as empirical results.}
\label{fig:provenance}
\end{figure}

Figure~\ref{fig:provenance} provides the legacy metric-provenance view; the V2 provenance record itself is machine-readable in \texttt{results/v2/provenance\_demo.json}.

\subsection{Enforcement-latency benchmark}

The current benchmark evaluates 500 trials per population size and includes only trials in which a violation was detected.

\begin{table}[htbp]
\centering
\caption{In-process enforcement latency conditional on detected violation. Values are means; the number of violating trials varies because only detected violations enter the timing distribution.}
\label{tab:latency}
\begin{tabular}{@{}rrrrr@{}}
\toprule
Operators & Attempted & Violating & Mean detect (ms) & Mean total (ms)\\
\midrule
1 & 500 & 55 & 0.019 & 0.028\\
10 & 500 & 42 & 0.061 & 0.072\\
50 & 500 & 48 & 0.220 & 0.234\\
200 & 500 & 53 & 0.697 & 0.715\\
500 & 500 & 59 & 1.650 & 1.682\\
\bottomrule
\end{tabular}
\end{table}

\begin{figure}[htbp]
\centering
\includegraphics[width=0.90\linewidth]{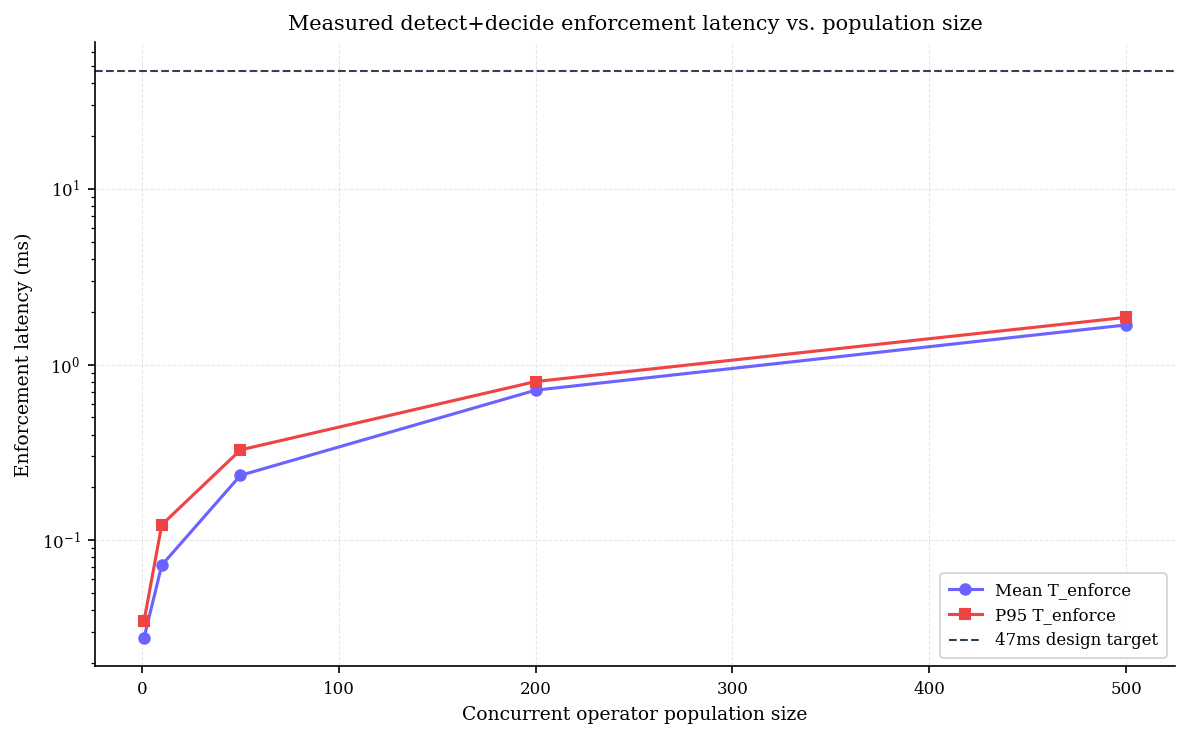}
\caption{Measured in-process detection-plus-decision latency conditional on a detected violation. The benchmark uses a single constructed SAT token and does not represent live network or regulator-infrastructure latency.}
\label{fig:latency}
\end{figure}

The approximately 60-fold increase from one to 500 operators is consistent with the fact that the AIB detection stage scales with the candidate population, while the decision stage remains comparatively small. These are software execution measurements under the stated environment and cannot be generalized to an operational spectrum-management platform without additional system-level measurements.

\subsection{Policy conformance and mutation testing}

Seven policy mutations were tested across protection-threshold, frequency-boundary, geographic-relocation, and administrative-metadata categories. All seven conformed to their predefined directional expectations, yielding a 100\% policy-conformance rate for this experiment with zero surviving mutants.

The result is useful as a regression of the decision path, but its scope is narrow. QoS-tier effects, jurisdictional routing, temporal validity, and conflicting-rule interactions were outside this experiment.

\section{What the Results Establish}
\label{sec:evidence}

The evidence can be separated into four levels.

\begin{table}[htbp]
\centering
\small
\caption{Claims supported by the current computational evidence.}
\label{tab:evidence-supported}
\begin{tabularx}{\linewidth}{@{}X l X@{}}
\toprule
Claim & Status & Evidence\\
\midrule
Regulatory constraints can be represented as executable computational rules & Demonstrated & Reference implementation and policy objects\\
Fixed-input compliance behaviour is reproducible & Demonstrated & Tests, deterministic seeds, and regenerated audit outputs\\
Per-operator computational decisions can carry causal provenance & Demonstrated & Selective authorization records and provenance demonstration\\
The engine can detect a deliberately constructed violation & Mechanism-level demonstration & Corrected co-located validation case\\
Controlled contention can reduce protection under the reference fixture & Simulator measurement & S0--S6 stress test\\
Selective authorization can preserve protection while allowing substantial access & Simulator measurement & Paired baseline and policy frontier\\
\bottomrule
\end{tabularx}
\end{table}

\begin{table}[htbp]
\centering
\small
\caption{Claims deliberately not established by the current study.}
\label{tab:evidence-not-established}
\begin{tabularx}{\linewidth}{@{}X l X@{}}
\toprule
Claim & Status & Reason\\
\midrule
Selective authorization is universally optimal, fair, or strategy-proof & \textbf{Not established} & The implemented policy is greedy\\
V2 protection rates are real-world protection probabilities & \textbf{Not established} & Scenarios and incumbent thresholds are controlled reference fixtures\\
Measured latency is end-to-end operational enforcement latency & \textbf{Not established} & Benchmark is in-process and conditional on detected violation\\
The \$4.0T historical quantity is realized economic benefit & \textbf{Not established} & Historical illustrative calibration, not an independently estimated market outcome\\
\bottomrule
\end{tabularx}
\end{table}

This distinction is not merely a reporting convention. It determines what can legitimately be transferred from the computational experiment to policy conclusions.

\section{Discussion: Policy Implications}
\label{sec:discussion}

\subsection{Machine-readable governance as an institutional interface}

The principal policy implication is that machine-readable spectrum governance can be treated as an institutional interface rather than merely a software implementation detail. A regulator's rule can be represented by explicit parameters for protected frequency, interference threshold, geographic exclusion, authority, and temporal validity. A computational system can then execute that rule consistently and retain evidence of which rule version was applied.

The value of this approach is strongest when rules change frequently. A static license may encode a fixed right, but an adaptive authorization system can recompute eligibility when the protected environment changes. The V2 policy frontier demonstrates the computational feasibility of this parameter sensitivity, but it does not establish that a particular threshold is socially optimal.

\subsection{Auditability requires more than reproducibility}

A reproducible simulation can show that the same input produces the same result. A governance system additionally needs to explain why a particular operator was authorized or suspended.

The V2 provenance model makes this distinction explicit. It records the operator's contribution, aggregate violation, effective threshold, standalone compliance status, suspension status, and geographic exclusion. In the demonstration, the deliberately constructed aggressor is suspended while a harmless background operator is authorized. This is a stronger computational explanation than simply reporting that an aggregate incumbent was violated.

The architecture therefore treats auditability as a first-class output of the authorization process rather than as a post-hoc logging step.

\subsection{Selective authorization exposes the protection--access trade-off}

The baseline experiment shows why the semantics of an adaptive authorization rule matter. Population-wide aggregate gating is simple and conservative, but it can make a local violation affect operators that did not materially contribute to the violation. Selective authorization preserves the same aggregate interference constraint while allowing the system to remove selected contributors until the constraint is restored.

The V2 benchmark shows a clear trade-off: selective authorization achieves 100\% protection and 89.43\% access opportunity in S3, whereas aggregate dynamic gating achieves 99.91\% protection and 99.53\% access. The appropriate policy choice depends on the regulator's objective function. If protection is lexicographically dominant, selective suspension is attractive; if maximizing access opportunity is prioritized subject to a small permitted protection risk, aggregate gating may be preferred.

Crucially, the identical results of the selective Dynamic-SAS and SPECTRA-GOV variants prevent us from attributing this trade-off to a unique SPECTRA-GOV admission algorithm. The contribution is the ability to place such a policy inside a broader governance and provenance architecture.

\subsection{Cross-jurisdiction coordination remains institutional}

GITR can provide a common computational representation of protected incumbents, but a shared data structure does not itself create legal authority. Questions of who may modify an incumbent record, how conflicting national rules are reconciled, how cross-border disputes are resolved, and how affected communities participate remain institutional questions.

The architecture therefore should be understood as a potential computational substrate for coordination rather than as a claim that coordination has already been solved. This is especially relevant to equity questions around orbital resources and developing-country access \citep{li2025activating}.

\subsection{Regulatory experimentation}

The DTRS concept and V2 policy frontier together suggest a practical regulatory-sandbox workflow. A regulator could instantiate a proposed rule, replay a fixed physical population, evaluate the resulting protection and access frontier, inspect causal provenance, and compare alternative policy semantics before deployment.

The methodological requirement is that the sandbox preserve the distinction between a controlled experiment and a prediction. A policy frontier generated from a synthetic population can reveal implementation consequences and internal trade-offs; it cannot, by itself, establish social welfare, fairness, or field performance.

\subsection{What the V2 audit changes about the research claim}

The most important outcome of the V2 work is therefore not a larger headline percentage. It is a more precise boundary around the contribution.

The audit identified and corrected a number of issues that would otherwise have weakened the evidentiary chain: invalid mixed-unit distance calculation, unintended spectral overlap in the control class, non-deterministic class seeding across processes, a PIEZ safety-margin mismatch, missing geographic exclusion in selective authorization, sub-MHz overlap errors, mutable GITR history, mixed simulated and wall-clock time, and an incorrectly paired static baseline. Correcting these issues does not prove the architecture. It makes the computational evidence more defensible by ensuring that the experiment measures what its labels say it measures.

\section{Scope and Limitations}
\label{sec:limitations}

\subsection{Propagation model}

The propagation implementation is simplified and informed by ITU-R P.452 and P.619 concepts; it is not a complete regulatory-grade implementation of either Recommendation. The corrected geodesic distance removes the most direct mixed-unit error from the V1 calculation, but it does not make the overall propagation model empirically validated.

\subsection{Reference incumbent thresholds}

The $I_{\max}$ values in the GITR fixture are reference calibration values. The current experiments therefore establish behaviour relative to the configured thresholds, not protection guarantees for real radar, railway, AIS, or aviation systems.

\subsection{Scenario construction}

The V2 scenario generator deliberately constructs overlap and proximity. This is an improvement over the V1 scenario configuration for mechanism evaluation, but it remains synthetic. The S4 adversarial class is intentionally severe and should not be interpreted as a statistical model of malicious or accidental interference.

S6 is also deliberately limited: it varies the active population between scenario draws rather than implementing a true within-scenario time series of operators joining and leaving.

\subsection{Selective authorization}

The V2 selective mechanism is a greedy largest-contributor-first rule. It is not proven optimal, fair, incentive compatible, or strategy-proof. Alternative policies could incorporate operator value, fairness, QoS, jurisdiction, priority, or historical obligations and could produce different trade-offs.

\subsection{Governance completeness}

The current conformance experiment does not fully exercise QoS tiers, jurisdictional interactions, temporal validity, or conflicting rules across multiple authorities. These are essential to a full regulatory implementation.

\subsection{Enforcement latency}

The V2 latency benchmark is in-process Python wall-clock timing conditional on a detected violation and using a single SAT token. It excludes network transport, cryptographic communication, external databases, monitoring hardware, regulator infrastructure, and human procedural latency. It therefore should not be compared directly with an end-to-end regulatory enforcement service.

\subsection{Economic quantities}

The original 4.4$\times$ spectrum-efficiency quantity and \$4.0T economic value remain historical illustrative/calibrated quantities. The V2 experiments do not independently validate them. No economic conclusion should be inferred from the V2 access-opportunity percentage alone.

\subsection{Accountability and affected communities}

Computational provenance is not equivalent to social accountability. The present work can show what rule and data produced a computational decision; it does not establish that affected communities can contest the decision, that outcomes are equitable, or that the governance architecture has legitimate institutional authority.

\section{Reproducibility and Research Integrity}
\label{sec:reproducibility}

The complete implementation, tests, audit scripts, generated evidence, and reproducibility instructions are maintained in the accompanying repository. The current V2.3 checkpoint was verified with 196 passing automated tests, successful Python compilation, and a clean whitespace check. The V1 regression was rerun from the current tree and reproduced the original as-shipped aggregate outputs.

The V2 evidence snapshot records:

\begin{itemize}
\item 10,000 scenarios per S0--S6 contention class;
\item 3,000 scenarios per class and 2,000 bootstrap resamples for uncertainty analysis;
\item 2,000 paired scenarios for the primary baseline comparison;
\item 1,000 scenarios per class for the all-class baseline comparison;
\item 600 scenarios for RCIA and S7 policy-change experiments;
\item 2,000 scenarios for the policy frontier;
\item 500 timing trials per population size for the enforcement benchmark.
\end{itemize}

The machine-readable outputs are the source of record for the reported V2 values. \texttt{RESULTS\_V2.md} is retained as historical revision/provenance material, while \texttt{RESULTS\_V2\_CURRENT.md} is the current evidence snapshot.

No V1 result was silently rewritten to make the V2 evidence appear more favourable. Corrections are represented as incremental versioned changes, and the original paper implementation remains separately identifiable at \texttt{v0.1.0-paper}.

\section{Conclusion}
\label{sec:conclusion}

This paper presented SPECTRA-GOV, a Tri-Layer Adaptive Governance Architecture for translating spectrum-governance constraints into computational authorization, protection, enforcement, and audit processes. The central contribution is an integrated governance architecture rather than a claim that any individual propagation model, blockchain mechanism, or selective admission heuristic is novel in isolation.

The V2 evaluation extends the original paper implementation without replacing it. The corrected code uses geodesic distance, explicit interval overlap, deterministic scenario seeds, corrected PIEZ safety margins and geographic exclusion, immutable registry updates, consistent simulated-time handling, paired baseline populations, and machine-readable provenance. The resulting experiments demonstrate that the implementation can operate under genuine controlled contention and can distinguish a deliberately constructed violation from a compliant case.

The quantitative results establish several conditional findings. Under the controlled S0--S6 scenarios, protection remained near 100\% in most classes and fell to 90.00\% in the deliberately adversarial close-proximity class. Selective authorization restored 100\% incumbent protection in the primary S3 baseline while retaining 89.43\% access opportunity, compared with 99.91\% protection and 99.53\% access under population-wide dynamic gating. The equality between the selective Dynamic-SAS and SPECTRA-GOV selective variants is an important boundary on the novelty claim: selective admission itself is not presented as uniquely novel. Instead, the distinctive contribution is the integration of policy execution, provenance, adaptive authorization, and cross-layer governance.

The study does not establish real-world protection probability, regulatory-grade propagation accuracy, fairness, strategic robustness, field enforcement latency, economic benefit, or institutional adoption. These are not minor qualifications; they define the next research agenda. Future work should replace the simplified propagation model with validated regulatory-grade calculations, introduce empirically grounded geographic and traffic distributions, model within-scenario temporal dynamics, evaluate strategic operator behaviour, exercise jurisdictional and conflicting-rule interactions, and validate the governance and accountability model with regulators and affected stakeholders.

SPECTRA-GOV is therefore best understood as a computational governance reference architecture and an auditable research platform. Its principal value is not that it produces a preferred protection percentage, but that it makes the chain from regulatory rule to computational decision explicit enough to be tested, corrected, reproduced, and challenged.

\section*{Declaration of competing interest}

The authors declare that they have no known competing financial interests or personal relationships that could have appeared to influence the work reported in this paper.

\section*{Data and code availability}

The complete reference implementation, tests, audit procedures, machine-readable V2 results, and reproducibility documentation are available in the project repository. The research-development history is preserved through versioned Git commits and tags. The original paper implementation is identified by \texttt{v0.1.0-paper}; the current post-audit checkpoint is identified by \texttt{v0.2.3-post-audit-hardening}. The machine-readable V2 evidence is contained in \texttt{results/v2/}, with the current evidence summary in \texttt{RESULTS\_V2\_CURRENT.md}.


\end{document}